\documentclass[aps,prl,twocolumn,showpacs]{revtex4}

\usepackage{graphicx}

\begin{document}

\title{Mixed-State Dissipation in Zero Temperature Limit in $MgB_2$ Thin Films
}

\author{Y. Jia$^1$, Y. Huang$^1$, H. Yang$^1$, L. Shan$^1$, C. Ren$^1$,
C. G. Zhuang$^{2,3,4}$, Y. Cui$^2$, Qi Li$^2$, Z. K. Liu$^{3}$, X.
X. Xi$^{2,3}$, and H. H. Wen$^1$}

\affiliation{$^1$National Laboratory for Superconductivity,
Institute of Physics and National Laboratory for Condensed Matter
Physics, Chinese Academy of Sciences, P.~O.~Box 603, Beijing 100080,
P.~R.~China}

\affiliation{$^2$Department of Physics, The Pennsylvania State
University, University Park, Pennsylvania 16802, USA}

\affiliation{$^3$Department of Materials Science and Engineering,
The Pennsylvania State University, University Park, Pennsylvania
16802, USA}

\affiliation{$^4$Department of Physics, Peking University, Beijing
100871, PR China}

\date{\today}

\begin{abstract}

We have studied mixed-state dissipation in epitaxial MgB$_2$ thin
films by measurements of resistive transition, current-voltage
characteristics, Hall effect, and point-contact tunnelling spectrum.
We found that unlike single gap superconductors with negligible
vortex quantum fluctuations in which vortices are frozen at $T=0$ K,
finite zero-temperature dissipation due to vortex motion exists in
MgB$_2$ over a wide magnetic field range. This dissipation was found
to be associated with proliferation of quasiparticles from the
$\pi$-band of MgB$_2$. The result shows that the vortex fluctuations
are enhanced by two-band superconductivity in MgB$_2$ and we suggest
that the vortex quantum fluctuation is a possible cause of the
non-vanishing zero-temperature dissipation.

\end{abstract}


\maketitle

Anomalous properties of vortex matter have been observed or
predicted in MgB$_2$ in association with two-gap superconductivity
\cite{Eskildsen,Koshelev,Gurevich,Vinokur,Bouquet}. In MgB$_2$ there
are 3-dimensional (3D) $\pi$ bands (the energy gap $\sim$ 2 meV) and
2-dimensional (2D) $\sigma$ bands (the energy gap $\sim$ 7 meV)
\cite{Choi,Iavarone}. The coherence length in the $\sigma$ band is
smaller than that in the $\pi$ band, in particular in the $c$
direction \cite{Koshelev2}. Thus the vortices in MgB$_2$ are 3D-like
and large with the $\pi$ gap \cite{Eskildsen} and 2D-like with
smaller cores if the $\pi$-band superconductivity is suppressed. It
has been observed that the resistive transition width increases
monotonically with increasing magnetic field in MgB$_2$
\cite{Tajima,Pradhan,JinH,Larbalestier}, or equivalently the
magnetic field transition width $\Delta H_c(T)=H_{c2}(T)-H_{irr}(T)$
increases monotonically with decreasing temperature \cite{Wen}. In a
type-II superconductor, $\Delta H_c(T)$ should vanish at $T=0$ K
because thermal activation of flux motion disappears, and
consequently there should be no mixed-state dissipation at $T=0$ K
\cite{Brandt}, unless strong quantum vortex fluctuations exist
\cite{Blatter}. In this paper we present results from high quality
epitaxial MgB$_2$ thin films, which show that the monotonic
broadening of the resistive transition leads to non-vanishing
mixed-state dissipation in the zero temperature limit and at
sufficiently high magnetic field that suppresses the $\pi$ band
superfluid density. $\Delta H_c(T)$ is over 7 T at $T=0$ K. Under
the high field, the vortex system changes from 3D-like to 2D-like
with smaller vortex cores, and the spread of the $\pi$-band
quasiparticles reduces the entropy difference inside and outside of
the $\sigma$ band-dominated vortex cores, enhancing vortex
fluctuations in MgB$_2$. We suggest that vortex quantum fluctuation
is a possible cause for the zero-temperature dissipation, made
observable in MgB$_2$ by the two-gap superconductivity.

The MgB$_2$ thin films used in this study were grown by the hybrid
physical-chemical-vapor deposition technique \cite{ZengXH} on (0001)
4H-SiC substrate. They are epitaxial with the $c$ axis normal to the
film surface. The films, with a thickness of about 100 nm, were
patterned into bridges with a width of 20 $ \mu $m for the resistive
measurement, and into Hall bars with a width of 0.4 mm for Hall and
point-contact tunnelling measurements. All measurements were made on
an Oxford cryogenic Maglab-Exa-12 system with magnetic field up to
12 T. In this paper, results obtained from one MgB$_2$ thin film are
presented, but similar effect was also observed in other films.

\begin{figure}
\includegraphics[width=8cm]{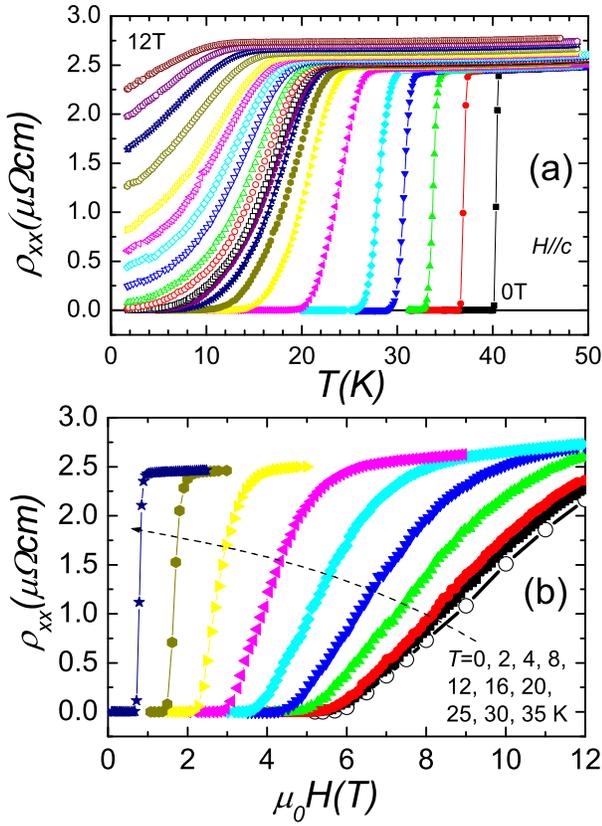}
\caption{(Color online) (a) Temperature dependence of resistive
transitions at various magnetic fields (right to left): 0, 0.5, 1,
1.5, 2, 3, 4, 4.5, 5, 5.2, 5.4, 5.7, 6, 6.5, 7, 7.5, 8, 9, 10, 11,
12 T. Above about 5.2 T zero resistivity is not reached in the
zero-temperature limit. (b) Magnetic field dependence of resistivity
at temperatures of (right to left) 0, 2, 4, 8, 12, 16, 20, 25, 30,
35 K. The $T=0$ K data were extracted from the data between $T=1.7$
K and $T=3$ K using a linear fit. A broad transition over 7 T is
observed at $T=0$ K.} \label{fig1rth}
\end{figure}

\begin{figure}
\includegraphics[width=8cm]{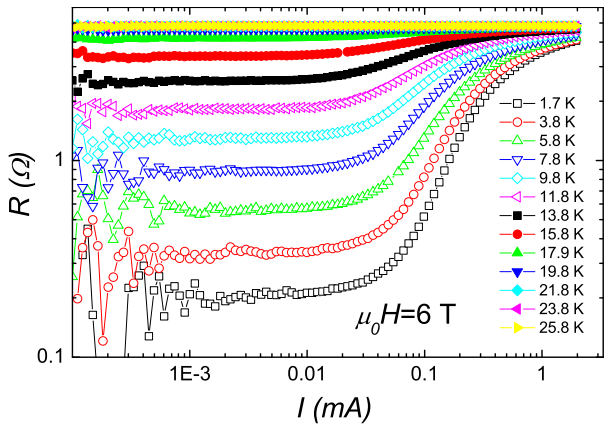}
\caption{(Color online) $I-V$  characteristics ($R=V/I$) at
different temperatures under a field of $\mu_0H=6$T. At the lowest
temperature $T=1.7$ K, the resistivity is finite in the zero current
limit indicating an equilibrium state property. } \label{fig2iv}
\end{figure}

In Fig. 1(a) we present the temperature dependence of resistivity
measured with the magnetic field normal to the film surface. At zero
field, the onset transition temperature is 40.74 K with a width
$\Delta T_c\sim 0.46$ K (determined with $1\%-99\%\rho_n$), and the
normal state residual resistivity $\rho_n$ is about 2.4 $\mu \Omega
cm$. When the magnetic field increases, $T_c$ decreases and the
transition is broadened monotonically. At fields above 5.2 T, zero
resistivity is not achieved at any temperature, suggesting a
non-vanishing dissipation in the zero-temperature limit. In Fig.
1(b) we present the field dependence of resistivity at various
temperatures. Instead of becoming narrower when temperature
approaches zero as in single gap superconductors with negligible
vortex quantum fluctuations \cite{Brandt}, the transition width
$\Delta H_c(T)$ increases monotonically when the temperature is
decreased down to 1.7 K, the lowest temperature in our measurements.
A zero temperature resistivity at different fields is extracted by
using a linear fit to the data between 1.7 K and 3 K, which is
plotted as open circles in Fig. 1(b). The extracted data at $T = 0$
K show finite resistivity over a broad magnetic field range ($>$ 7
T). To make sure that the finite dissipation at $T = 0$ K is an
equilibrium state property and not the result of a large driving
force due to measurement current, we measured $I-V$ curves at
different magnetic fields and an example is shown in Fig. 2 for
$\mu_0H=6$ T. It is clear that finite dissipation in the low
temperature limit at fields above 5.2 T is obtained in the zero
current limit.

It is necessary to determine the nature of the zero-temperature
dissipation at high fields ($\mu_0 H\geq$ 5.2 T). It has been
pointed out by Vinokur {\it et al.} \cite{VinokurHall} that there is
a general scaling law for dissipations due to vortex motion:
$\rho_{xy}=A \rho_{xx}^\beta$ with $\beta$ close to 2, where
$\rho_{xy}$ and $\rho_{xx}$ are Hall and longitudinal resistivity,
respectively, and $A$ is a field dependent parameter. In Fig. 3, we
show the relationship between $\rho_{xy}$ and $\rho_{xx}$ during the
resistive transition in a log-log plot for various fields. The
decreasing $\rho_{xx}$ value in a curve corresponds to decreasing
temperature. A scaling relationship with $\beta=2$ was found for all
the fields, indicating that the nature of the dissipation in the
MgB$_2$ film is vortex motion. The fitted $\beta$ value in the low
temperature region ($\rho_{xx}\leq 1/3\rho_n$, where $\rho_n$ is the
normal state resistivity) is plotted against the field and shown in
the inset (a), and $\beta=2.0\pm0.1$ is obtained. A typical example
for fitting with upper $T \rightarrow$ 0 is shown in inset (b) for
$\mu_0 H =$ 6 T. It shows that $\beta$ is close to 2 at low
temperatures but deviates slightly from 2 at around 10 K. It
suggests that in the zero temperature limit the dissipation arises
from vortex motion. Above $\rho_{xx} \sim 1/3\rho_n$ (about 15 K for
6 T), the contribution of the qusiparticle scattering cannot be
neglected. Similar result has been reported by Kang {\it et al.}
\cite{KangScaling}.

\begin{figure}
\includegraphics[width=8cm] {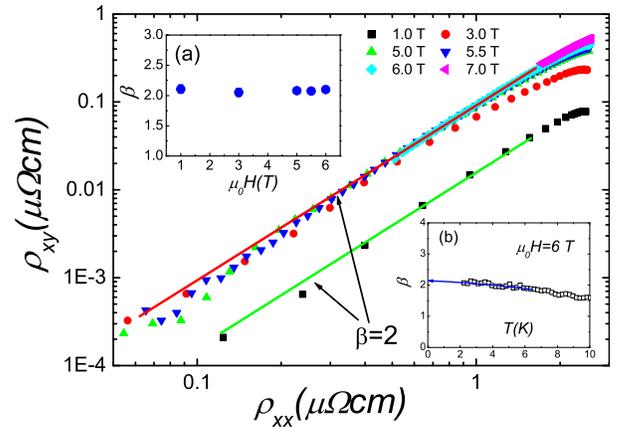}
\caption{(Color online) Correlation between $\rho_{xx}$ and
$\rho_{xy}$ measured at different magnetic fields. The two solid
lines indicate the slope $\beta=2$. Inset (a) The field dependence
of $\beta$ determined in the region $\rho_{xx}\leq 1/3\rho_n$. Inset
(b) Dependence of $\beta$ on the upper limit of $T$ during fitting.
} \label{fig3hall}
\end{figure}

In the earlier reports of transition broadening in polycrystalline
MgB$_2$ samples \cite{Pradhan,JinH}, there is a possibility that it
may be caused by the different upper critical fields for $H||c$ and
$H||ab$ as the orientations of the grains are random. The single
crystal-like epitaxial MgB$_2$ films used in this work allow
measurements for $H||c$ and $H||ab$ separately. In Fig. 4, the upper
critical field and the irreversibility field for the MgB$_2$ film
are shown for both $H||c$ and $H||ab$. The values for $H_{c2}$ and
$H_{irr}$ are determined using the criteria of $\rho_{xx}=99\%
\rho_n$ and $\rho_{xx}=1\times 10^{-3}\mu \Omega$$cm$, respectively.
The figure shows a monotonic increase of $\Delta H_c(T)$ with
decreasing temperature and a large $\Delta H_c(T)$ over 7 T at $T=$
0 K for $H||c$. The $\Delta H_c(T)$ at $T=$ 0 K observed here is
much larger than those in the previous reports of vortex quantum
fluctuations \cite{Blatter,Okuma}. The figure also shows that
percentage-wise the broadening of $\Delta H_c(T)$ is much larger for
$H||c$ than for $H||ab$. The critical field from the tunnelling
spectrum (see Fig. 5 and related text) is also presented.

\begin{figure}
\includegraphics[width=8cm]{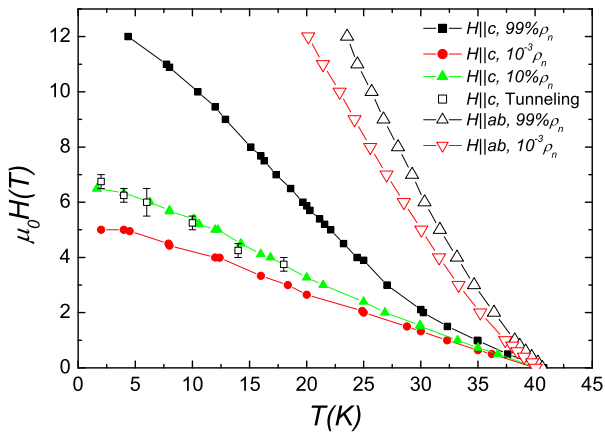}
\caption{(Color online) The vortex phase diagram determined using
different criterions/techniques. The fields at which the $\pi$-band
gap feature in the point-contact tunnelling spectra disappears are
also shown as open squares.}\label{fig4phase}
\end{figure}

Using point-contact spectroscopy, we have found a strong correlation
between the non-vanishing zero temperature dissipation and the
suppression of the $\pi$-band superfluid density and subsequently
the proliferation of the quasiparticles from the $\pi$-band. In Fig.
5(a), the point-contact tunnelling spectra at 2 K with the tip along
the $c$ axis are shown. The sharp Andreev reflection peaks in the
zero field spectrum correspond to the $\pi$-band gap and the slight
shoulders correspond to the $\sigma$-band gap. By fitting the
spectra to the extended BTK model \cite{BTK} using a linear
combination of the two contributions from the $\pi$ band and
$\sigma$ band, we found that the zero-field value is 2 mV for the
$\pi$ gap and 6.65 mV for the $\sigma$ gap, in good agreement with
previous experimental results \cite{Yanson,Caplin}. The dominant
contribution to the spectra is from the $\pi$-band, with a ratio of
the $\sigma$-band to $\pi$-band contributions of 0.075, indicating a
well-defined $c$-axis injection of quasiparticles in the tunnelling
measurement. The $\pi$-band Andreev reflection peaks decay rapidly
with increasing field and disappear completely at about 6.5 T. This
field is close to the field above which zero resistivity is not
obtained at $T=$ 0 K. In fact, as shown in Fig. 4, the field where
the Andreev peaks from the $\pi$-band is completely suppressed
overlaps with the temperature dependence of the magnetic field when
$\rho_{xx}=10\% \rho_n$.

From the suppression of the Andreev reflection peaks we have
estimated the quasiparticle density of state (DOS), $S_{qp}$, as a
function of the applied field. All the spectra were first normalized
by the conductance at 10 mV. Then $S_{qp}(H)$ was calculated using a
relation $S_{qp}(H) = I(0)-I(H)$, where $I(H)$ is the integrated
area under the spectrum between $\pm 10mV$, which is roughly
proportional to the superfluid density \cite{Yanson}, and
$I(0)=2.85$. The result is presented in Fig. 5(b) for field up to 2
T, as the error of calculation becomes too large at high fields. It
shows that the quasiparticle DOS increases rapidly and reaches about
67\% of the maximum value (maximum $S_{qp}$=2.85 at 6.5 T) at 2 T.
Also plotted in Fig. 5(b) is the normalized zero-bias conductance
(ZBC) from the STM measurement by Eskildsen {\it et al.} taken in
the superconducting region between two vortices \cite{Eskildsen},
which was shown to be consistent with the DOS determined from the
specific heat measurement \cite{Junod}. The calculated $S_{qp}(H)$
and the normalized ZBC from the STM measurement overlap remarkably
well. The fact that $S_{qp}(H)$ reaches only 67\% rather than near
90\% as in the STM measurement may be due to the normal cores of
vortices, which also contribute to the conductance in the
point-contact tunneling measurement. The result clearly shows that
the depression of the $\pi$-band Andreev peaks is accompanied by a
rapid increase of the quasiparticle DOS in the MgB$_2$ film.

\begin{figure}
\includegraphics[width=8cm]{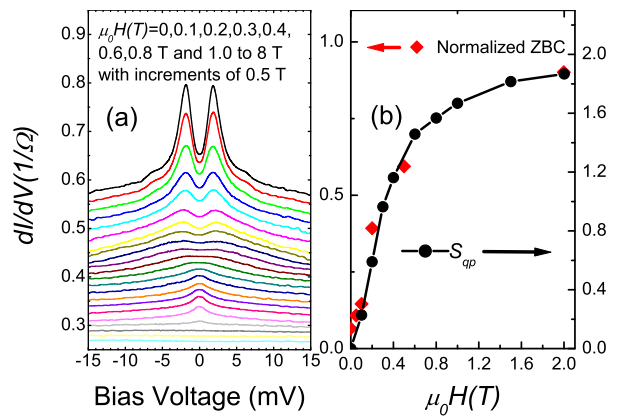}
\caption {(Color online) (a) Point-contact tunnelling spectra
measured at $T = 2$ K with magnetic fields from $H=0$ T (top) to 8 T
(bottom). Curves above $H=0$ T are shifted for clarity. The feature
due to the $\pi$-band gap is suppressed completely above 6.5 T. (b)
The calculated quasiparticle DOS (filled circles) derived from the
tunneling data. Normalized zero-bias conductance (ZBC) derived from
the STM measurement by Eskildsen {\it et al.} is also shown
\cite{Eskildsen}. } \label{fig5tun}
\end{figure}

Based on the results presented above, we can conclude that the
broadening of the resistive transition and the zero-temperature
dissipation in high magnetic field are the results of enhanced
vortex fluctuations in MgB$_2$ due to two-band superconductivity. In
MgB$_2$ the in-plane coherence length is about 51 nm for the $\pi$
band and 13 nm for the $\sigma$ band \cite{Eskildsen}. In the $c$
direction, the coherence length of the $\sigma$ band is 1/17 of that
of the $\pi$ band \cite{Koshelev2}. At low magnetic field, the
effective in-plane coherence length and thus the vortex core size is
dominated mainly by the $\pi$-band gap, and the vortices are thick,
3D and continuous. When the magnetic field is large enough to
suppress the $\pi$-band gap, the superfluid in MgB$_2$ is
predominantly from the $\sigma$-band and the vortices become 2D and
pancake-like, and the vortex core size shrinks. In addition, the
$\pi$-band quasiparticles are widespread in MgB$_2$
\cite{Eskildsen}, which reduces the entropy difference inside and
outside of the vortex core leading to smaller pinning energy at high
field as reported in the literature \cite{Wen,JChen}. These effects
cause strong vortex fluctuations in MgB$_2$ in high fields. When
$H||ab$, the broadening is less pronounced because the vortices are
always continuous. For practical applications, dopant and defect
need to be introduced into MgB$_2$, which increases interband and
intraband scattering, lowers anisotropy, and enhances vortex pinning
\cite{Braccini,Wen,JChen}.

Although the field-induced broadening of the resistive transition
above 5 K was interpreted as the result of thermally activated flux
motion \cite{Larbalestier}, while it is known that the thermally
activated flux motion should be absent at $T=0$ K. A possible
explanation for the zero-temperature dissipation is the vortex
quantum fluctuation, or the zero-point motion of vortices.
Traditionally, strong quantum fluctuations occur in 2D vortex
systems with large sheet resistance and small coherence length, that
is, the ratio $\rho_n/\xi$ needs to be large \cite{Blatter,Okuma}.
For the MgB$_2$ samples studied here, $\rho_n$ is small. However,
unlike single-gap $s$-wave superconductors where quasiparticles are
confined within the vortex cores at $T=0$ K, in MgB$_2$ the
$\pi$-band quasiparticles are both inside and outside of the vortex
cores \cite{Eskildsen}, which could have the same effect as disorder
in single-gap superconductors which enhances vortex quantum
fluctuation \cite{Blatter}. On the other hand, quasiparticles in the
superconductor could increase vortex viscosity which is detrimental
to quantum fluctuation. However, compared to the cuprate
superconductor, a 2D vortex system with weak pinning suggested for
strong vortex quantum fluctuation \cite{Blatter}, the vortex
viscosity characterized by the Bardeen-Stephen coefficient
$\eta\propto/\rho_n \xi^2$ in MgB$_2$ is comparable as the smaller
$\rho_n$ is balanced by the larger $\xi$. It is possible that the
two-band nature makes MgB$_2$ a suitable system to observe vortex
quantum fluctuation.

In summary, a non-vanishing mixed state dissipation has been
observed in MgB$_2$ thin films at $T=0$ K when the magnetic field is
high enough to suppress the $\pi$-band gap. Hall effect measurement
confirms that the zero-temperature dissipation is induced by vortex
motion. Point-contact tunneling spectroscopy indicates that it is
associated with the proliferation of quasiparticles from the $\pi$
band. The transition from a $\pi$ band-dominated 3D vortex system to
a $\sigma$ band-dominated 2D vortex system at high field and the
reduction of pinning energy due to the spread of $\pi$-band
quasiparticles inside and outside of the vortex core cause enhanced
vortex fluctuations in MgB$_2$. Vortex quantum fluctuation is a
possible mechanism for the dissipation at $T=0$ K.

We thank Alex Gurevich for helpful discussions. This work is
supported by the National Science Foundation of China, the Ministry
of Science and Technology of China (the 973 Project: 2006CB01000 and
2006CB921802), and the Chinese Academy of Sciences project (ITSNEM).
The work at Penn State is supported by NSF under grant Nos.
DMR-0306746 (XXX), DMR-0405502 (QL), and DMR-0514592 (ZKL and XXX),
and by ONR under grant No. N00014-00-1-0294 (XXX).

Corresponding author hhwen@aphy.iphy.ac.cn

\end{document}